\RequirePackage{lineno}
\documentclass[twocolumn,showpacs,aps,prd]{revtex4-1}
\usepackage{amsmath}
\usepackage{amssymb}
\usepackage{graphicx}
\usepackage{dcolumn}
\usepackage{bm}
\usepackage{subfigure}
\usepackage{amsfonts}
\usepackage{keyval,graphicx}
\usepackage{textcomp,wasysym}
\usepackage{multirow}
\usepackage{xcolor}
\usepackage[dvipdfm,CJKbookmarks=true,unicode,colorlinks,linkcolor=blue,anchorcolor=blue,citecolor=blue,pdfborder={0 0 0}]{hyperref}
\newcommand{\br}[1]{\mathcal{B}#1}
\newcommand{\el}[1]{\mathcal{L}#1}

\newcommand\ST{\rule[-0.5em]{0pt}{1.5em}}

\begin{document}
\normalsize

\title{\boldmath Observation of pseudoscalar and tensor resonances in $J/\psi\rightarrow \gamma \phi \phi$}

\author{
  \small
      M.~Ablikim$^{1}$, M.~N.~Achasov$^{9,e}$, X.~C.~Ai$^{1}$,
      O.~Albayrak$^{5}$, M.~Albrecht$^{4}$, D.~J.~Ambrose$^{44}$,
      A.~Amoroso$^{49A,49C}$, F.~F.~An$^{1}$, Q.~An$^{46,a}$,
      J.~Z.~Bai$^{1}$
% ...
      \\                 
      \vspace{0.2cm}     
      (BESIII Collaboration)\\
      \vspace{0.2cm} {\it
        $^{1}$ Institute of High Energy Physics, Beijing 100049, People's Republic of China\\
% ...
$^{4}$ Bochum Ruhr-University, D-44780 Bochum, Germany\\
        $^{5}$ Carnegie Mellon University, Pittsburgh, Pennsylvania 15213, USA\\
% ...
$^{9}$ G.I. Budker Institute of Nuclear Physics SB RAS (BINP), Novosibirsk 630090, Russia\\
% ...
$^{44}$ University of Rochester, Rochester, New York 14627, USA\\
% ...
$^{46}$ University of Science and Technology of China, Hefei 230026, People's Republic of China\\
% ...
$^{49}$ (A)University of Turin, I-10125, Turin, Italy; (B)University of Eastern Piedmont, I-15121, Alessandria, Italy; (C)INFN, I-10125, Turin, Italy\\
% ...
$^{a}$ Also at State Key Laboratory of Particle Detection and Electronics, Beijing 100049, Hefei 230026, People's Republic of China\\
% ...
$^{e}$ Also at the Novosibirsk State University, Novosibirsk, 630090, Russia\\
% ...
      }
}


\begin{abstract}
  Based on a sample of $(1310.6 \pm 10.5) \times 10^{6}$ $J/\psi$ events collected with
  the BESIII detector operating at the BEPCII storage ring, a partial wave analysis of the
  decay $J/\psi\rightarrow \gamma \phi \phi$ is performed in order to study the
  intermediate states. Results of the partial wave analysis show that the structures are
  predominantly $0^{-+}$ states. The existence of the $\eta(2225)$ is
  confirmed, and its resonance parameters are measured. Two
  additional pseudoscalar states, the $\eta(2100)$ with a mass of
  $2050_{-24}^{+30}{}_{-26}^{+75}$~MeV/$c^{2}$ and a width of
  $250_{-30}^{+36}{}_{-164}^{+181}$~MeV/$c^{2}$ and the $X(2500)$ with a mass of
  $2470_{-19}^{+15}{}_{-23}^{+101}$~MeV/$c^{2}$ and a width of
  $230_{-35}^{+64}{}_{-33}^{+56}$~MeV/$c^{2}$, are observed. In addition to these three
  pseudoscalar states, the scalar state $f_0(2100)$, and three tensor states, the
  $f_2(2010)$, $f_2(2300)$ and $f_2(2340)$, are observed in the
  process $J/\psi\rightarrow \gamma \phi\phi$.  The product branching fractions
  ${\cal B}$($J/\psi\rightarrow \gamma X)\times{\cal B}(X\rightarrow \phi \phi)$ are reported.
\end{abstract}

\pacs{13.20.Gd, 13.66.Bc, 14.40.Be}

\maketitle

\section{Introduction}
In quantum chromodynamics (QCD)$-$gluons$-$the gauge bosons of the strong force, carry color
charge and thus can form bound states called glueballs~\cite{bibg1,bibg2,bibg3}. The
search for glueballs is an important field of research in hadron physics.
However, possible mixing of the pure glueball states with nearby $q\bar{q}$ nonet mesons
makes the identification of glueballs difficult in both experiment and theory. The
glueball spectrum has been predicted by Lattice QCD~\cite{bib1,bib2,bib3}, where the
lowest-lying glueballs are scalar (mass 1.5$-$1.7 GeV/$c^2$), tensor (mass 2.3$-$2.4
GeV/$c^2$), and pseudoscalar (mass 2.3$-$2.6 GeV/$c^2$).  Radiative decays of the
$J/\psi$ meson provide a gluon-rich environment and are therefore regarded as one of the
most promising hunting grounds for glueballs~\cite{bibjpsi1,bibjpsi2}.

Broad $J^{PC}=2^{++}$ structures around 2.3 GeV/$c^2$ decaying to $\phi\phi$ were reported in
$\pi^{-} N$ reactions~\cite{bibpiN,bibpiBe} and in $p\bar{p}$ central
collisions~\cite{bibcentral,bibjetset}. In Ref.~\cite{bibpiNcom1,bibpiNcom2}, a tensor
glueball was assumed to be mixed with conventional tensor resonances.  Aside from the
$\eta(2225)$, which was discovered in
$J/\psi\rightarrow\gamma \phi \phi$~\cite{bib2225dm2,bib2225mark3,bib2225bes2}, the
structures in the pseudoscalar sector above 2 GeV/$c^2$ are poorly understood.

In this paper, we present a partial wave analysis (PWA) of $J/\psi\rightarrow \gamma \phi \phi$,
where both $\phi$ mesons are reconstructed from $K^{+}K^{-}$, based on a
sample of $(1310.6 \pm 10.5) \times 10^{6}$ $J/\psi$ events collected with the BESIII
detector~\cite{bibbes3}.

\section{BESIII detector and Monte Carlo Simulation}
The BESIII detector is a magnetic spectrometer operating at BEPCII, a double-ring
$e^{+}e^{-}$ collider with center-of-mass energies between 2.0 and 4.6 GeV. The
cylindrical core of the BESIII detector consists of a helium-based main drift chamber
(MDC), a plastic scintillator time-of-flight system (TOF) and a CsI(Tl) electromagnetic
calorimeter (EMC) that are all enclosed in a superconducting solenoidal magnet providing a
magnetic field of 1.0 T (0.9 T in 2012, for about $1.09 \times 10^9$ $J/\psi$ events).  The
solenoid is supported by an octagonal flux-return yoke with resistive plate counter muon
identifier modules interleaved with steel. The acceptance for charged particles and photons
is 93\% of the 4${\pi}$ solid angle, and the charged-particle momentum resolution at $p$ =
1 GeV/$c$ is 0.5\%. The EMC measures photon energies with a resolution of 2.5\% (5\%) at
$E_{\gamma}$ = 1 GeV in the barrel (end caps).

A {\footnotesize GEANT}4-based~\cite{bibgeant4} Monte Carlo (MC) simulation software package is used to
optimize the event selection criteria, estimate backgrounds and determine the detection
efficiency. We generate a large signal MC sample of $J/\psi\rightarrow \gamma \phi \phi$,
$\phi\rightarrow K^+K^-$ uniformly in phase space.

\section{Event Selection}
Charged tracks in the polar angle range $|\cos \theta|<0.93$ are reconstructed from hits
in the MDC. The combined information from the energy
loss ($dE/dx$) measured in MDC and flight time in TOF is used to form particle identification
confidence levels for the $\pi$, $K$ and $p$ hypotheses. Each track is assigned the
particle type corresponding to the highest confidence level. Photon candidates are
required to have an energy deposition above 25 MeV in the barrel EMC ($|\cos \theta|< 0.80$)
and 50 MeV in the end cap EMC ($0.86 <|\cos \theta|< 0.92$).  To exclude showers from
charged particles, the angle between the shower position and the charged tracks
extrapolated to the EMC must be greater than 10 degrees. A requirement on the EMC timing
is used to suppress electronic noise and energy deposits unrelated to the event.

The study of the $\gamma K^{+} K^{-} K^{+} K^{-}$ final state is complicated by low
momentum kaons significantly affecting the reconstruction efficiency, especially at low
$\phi\phi$ masses. To improve the reconstruction efficiency, the
$J/\psi\rightarrow \gamma K^{+}K^{-}K^{+}K^{-}$ candidate decays are reconstructed with at
least one photon and at least three charged tracks identified as kaons. A one-constraint
(1C) kinematic fit under the hypothesis
$J/\psi\rightarrow \gamma K^{+}K^{-}K^{\pm}K^{\mp}_{\text{miss}}$ is performed by
constraining the mass of the missing particle to the kaon mass. The resulting
$\chi^{2}_{1C}$ is required to be less than 5. If more than one combination of one photon
and three kaon tracks meets this requirement, only the combination with the smallest
$\chi^2_{1C}$ is accepted.  To suppress possible background events with
$K^{+}K^{-}K^{+}K^{-}$ and $\pi^{0} K^{+}K^{-}K^{+}K^{-}$ final states, the $\chi^{2}$ of
a 1C kinematic fit under the hypothesis
$J/\psi\rightarrow K^{+}K^{-}K^{\pm}K^{\mp}_{\text{miss}}$ and the $\chi^{2}$ of a 2C
kinematic fit under the hypothesis
$J/\psi\rightarrow \pi^{0}K^{+}K^{-}K^{\pm}K^{\mp}_{\text{miss}}$, with an additional
constraint on the invariant mass of the two photons to be equal to the $\pi^{0}$ mass, are
both required to be larger than 10.

\begin{figure*}[!htbp]
  \centering
  \includegraphics[width=0.4\textwidth,height=0.24\textheight]{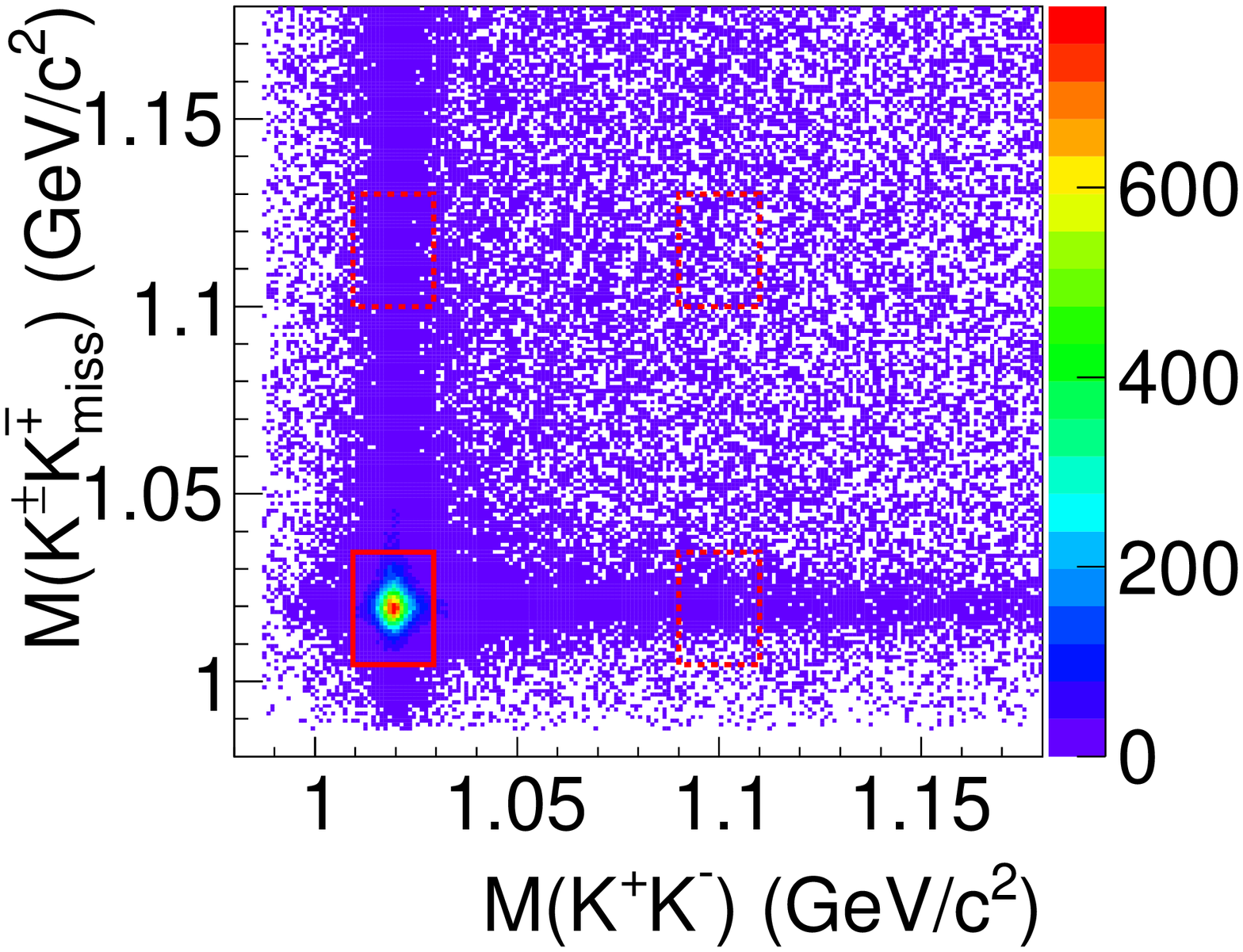}
  \put(-150,6){(a)}
  \includegraphics[width=0.4\textwidth,height=0.24\textheight]{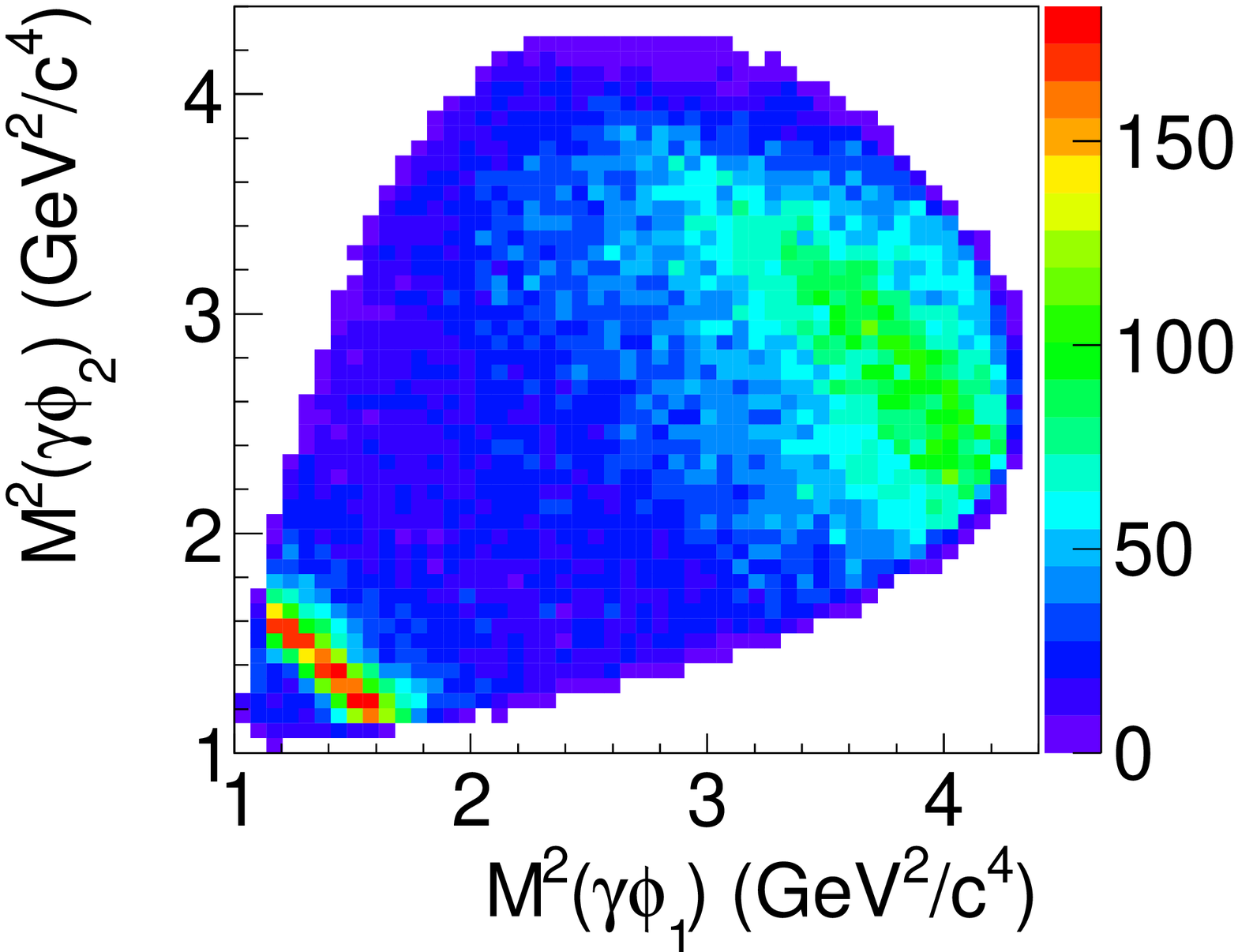}
  \put(-150,6){(b)}\\
  \includegraphics[width=0.4\textwidth,height=0.24\textheight]{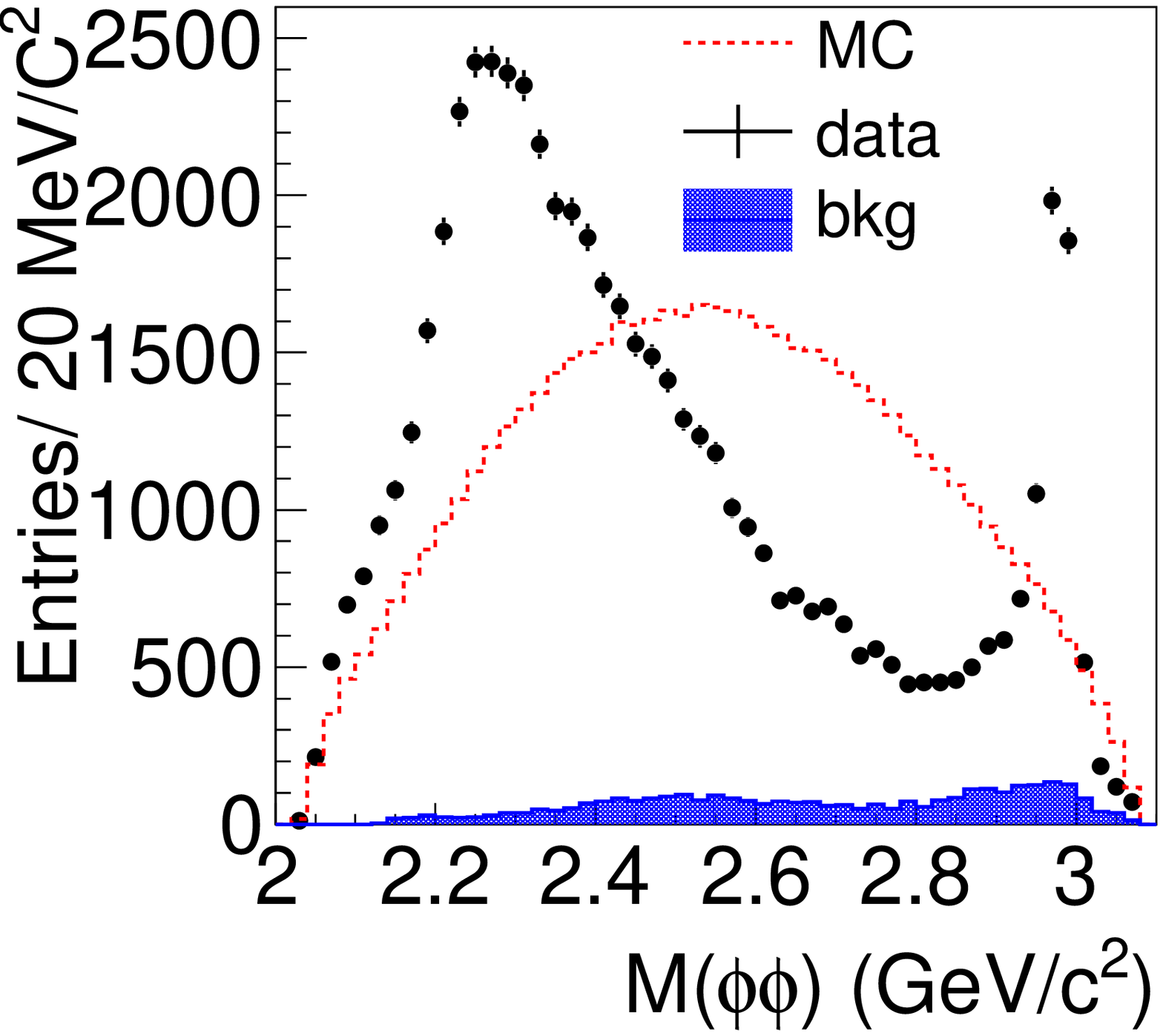}
  \put(-150,6){(c)}
   \includegraphics[width=0.4\textwidth,height=0.24\textheight]{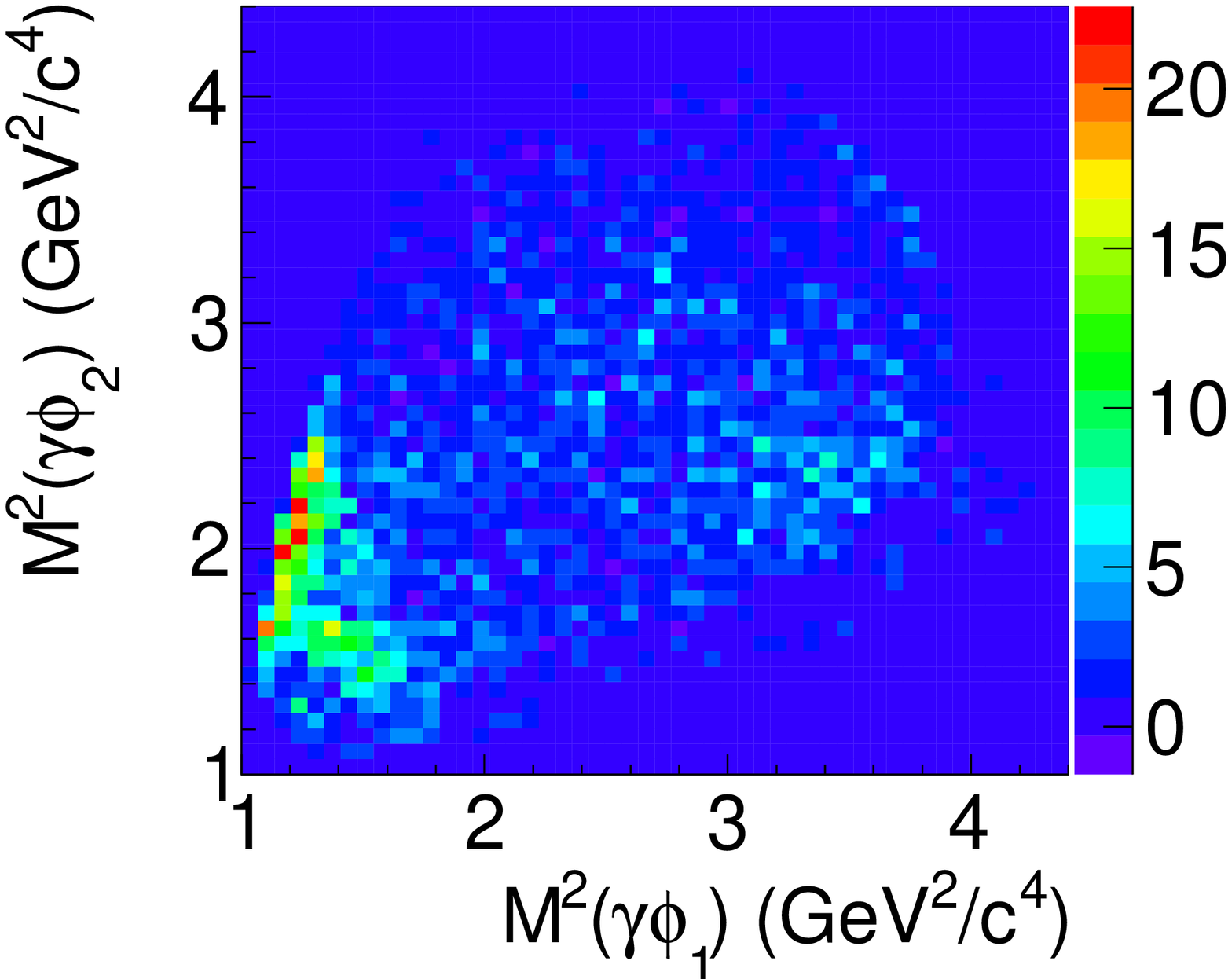}
  \put(-150,6){(d)}
  \caption{(a) Scatter plot of $M(K^{+}K^{-})$ versus
    $M(K^{\pm}K^{\mp}_{\text{miss}})$ for the selected $\gamma
    K^{+}K^{-}K^{\pm}K^{\mp}_{\text{miss}}$ candidates. The solid box and dashed
    boxes show the signal and sideband regions as defined in the text,
    respectively. (b) The corresponding Dalitz plot for the selected $\gamma\phi\phi$
    candidates. (c) Invariant mass distributions of
    $\phi\phi$ for the selected $\gamma\phi\phi$
    candidates. The points with error bars and the dashed line show data and simulation, respectively;
    the shaded histogram shows the background estimated from $\phi\phi$ sidebands. (d) The corresponding Dalitz plot for the background events estimated from $\phi\phi$ sidebands.}
  \label{fig:scatter}
\end{figure*}

For the selected $J/\psi\rightarrow\gamma K^{+}K^{-}K^{\pm}K^{\mp}_{\text{miss}}$
candidates, one $\phi$ is reconstructed from the $K^{+}K^{-}$ pair with an invariant mass
closest to the nominal pole mass $m_{\phi}$, and the other $\phi$ is reconstructed from
the remaining reconstructed kaon and the missing kaon. The scatter plot of $M(K^{+}K^{-})$
versus $M(K^{\pm}K^{\mp}_{\text{miss}})$ is shown in Fig.~\ref{fig:scatter}(a), where a
cluster of events corresponding to $\phi\phi$ production is evident. Because the processes
$J/\psi\rightarrow \phi\phi$ and $J/\psi\rightarrow \pi^{0}\phi\phi$ are forbidden by
$C$-parity conservation, the presence of two $\phi$ mesons is a clear signal for the
radiative decay $J/\psi \rightarrow \gamma\phi\phi$. The $\phi\phi$ events are selected by
requiring $|M(K^{+}K^{-}) - m_{\phi}| < 10$~MeV/$c^{2}$ (referred to as $\phi_1$) and
$|M(K^{\pm}K^{\mp}_{\text{miss}}) - m_{\phi}| < 15$~MeV/$c^2$ (referred to as $\phi_2$).
Simulation studies show that only 0.2\% of the selected $J/\psi \rightarrow \gamma\phi\phi$
events have a miscombination of kaons. The Dalitz plot and the invariant mass
distributions of $\phi\phi$ for the selected $\gamma\phi\phi$ candidate events are shown
in Fig.~\ref{fig:scatter}(b) and Fig.~\ref{fig:scatter}(c), respectively. A total of
58,049 events survive the event selection criteria. Besides a
distinct $\eta_c$ signal, clear structures in the $\phi\phi$ invariant mass spectrum are
observed.

Possible backgrounds are studied with a MC sample of $1.2\times10^{9}$
$J/\psi$ inclusive decays, in which the decays with known branching fractions are generated by
{\footnotesize EVTGEN}~\cite{bibevtgen} and the remaining $J/\psi$ decays are generated
according to the {\footnotesize LUNDCHARM}~\cite{biblundcharm1,biblundcharm2} model. The dominant backgrounds are found to be those with final states $\pi^{0}K^{+}K^{-}K^{+}K^{-}$,
$K^{+}K^{-}K^{\pm}\pi^{\mp}K_{L}$ and $\pi^{0}\pi^{0}K^{+}K^{-}K^{\pm}\pi^{\mp}$, such as
$J/\psi\rightarrow \phi f_1(1420)$, $f_1(1420)\rightarrow K\bar{K}\pi$ and $J/\psi\rightarrow\phi K^{*\pm} K^{\mp}$.  No
background event with $\phi\phi$ in the final state is observed.

Non-$\phi\phi$ backgrounds are estimated using the $\phi$ sideband events from data.
The two dimensional sidebands are illustrated by dashed boxes in
Fig.~\ref{fig:scatter}(a), where the sideband regions are defined as $1.09$~GeV$/c^{2} <
M(K^{+}K^{-}) < 1.11$~GeV/c$^{2}$ and $1.10$~GeV$/c^{2} < M(K^{\pm}K^{\mp}_{\text{miss}})
< 1.13$~GeV$/c^{2}$.  The shaded histogram
in Fig.~\ref{fig:scatter}(c) shows the background contribution estimated from the
normalized sideband events, corresponding to a background level of 5.4\%. The Dalitz plot for the estimated background events are shown in Fig.~\ref{fig:scatter}(d), where the accumulation of events in the left lower corner is mainly due to background events from $J/\psi\rightarrow \phi f_1(1420)$.

\section{Partial Wave Analysis}
\subsection{Analysis method}
Using the GPUPWA framework~\cite{bibgpu}, a PWA is performed on 45,852 events in the region
$M(\phi\phi) < 2.7$~GeV/$c^2$ in order to disentangle the structures present in the light mesons. Due to the detector resolution not being included in the PWA fit, events in the $\eta_c$ signal region are excluded. The quasi two-body decay amplitudes in the sequential decay process $J/\psi\rightarrow\gamma X$,
$X\rightarrow\phi\phi$, $\phi\rightarrow K^{+}K^{-}$ are constructed using the covariant
tensor amplitudes described in Ref.~\cite{bibcovformula}. $J/\psi\rightarrow \phi f_1(1285)$, $f_1(1285)\rightarrow \gamma \phi$ is ignored due to its low branching fraction~\cite{bibpdg}.
For the radiative $J/\psi$ decay to mesons, the general form for the decay amplitude is
\begin{equation}
A=\psi_{\mu}(m_{1})e_{\nu}^{*}(m_{2})A^{\mu\nu}=\psi_{\mu}(m_{1})e_{\nu}^{*}(m_{2})\Sigma_{i}\Lambda_{i}U_{i}^{\mu\nu},
\end{equation}
where $\psi_{\mu}(m_{1})$ is the $J/\psi$ polarization four-vector,
$e_{\nu}(m_{2})$ is the polarization vector of the photon and
$U_{i}^{\mu\nu}$ is the partial wave amplitude with coupling strength
determined by a complex parameter $\Lambda_{i}$. The partial wave
amplitudes $U_{i}$ used in the analysis
are constructed with the four-momenta of the particles in the final state, and their
specific expressions are given in Ref.~\cite{bibcovformula}.

In this analysis, we use Breit-Wigner (BW) as an approximation to describe the leading singularity since no model is available yet for the high energy region with many channels opened. Each resonance $X$ is
parametrized by a constant-width, relativistic BW propagator,
\begin{equation}
BW(s)=\frac{1}{M^{2}-s-iM\Gamma},
\end{equation}
where $s$ is the invariant mass-squared of $\phi\phi$, and $M$
and $\Gamma$ are the mass and width of the intermediate resonance.

The complex coefficients of the amplitudes and resonance parameters are determined by an
unbinned maximum likelihood fit with the likelihood function constructed as in Ref.~\cite{bibgee}.

The probability to observe the $i$th event characterized by
the measurement $\xi_i$, i.e., the measured four-momenta of the particles in the final state, is:
\begin{equation}
P(\xi_i)=\frac{\omega(\xi_i)\epsilon(\xi_i)}{\int d\xi\omega(\xi)\epsilon(\xi)},
\end{equation}
where $\epsilon(\xi_i)$ is the detection efficiency and
$\omega(\xi_i)\equiv(\frac{d\sigma}{d\Phi})_i$ is the differential cross
section, and $d\Phi$ is the standard element of phase space. The full
differential cross section is:
\begin{equation}
\small
\frac{d\sigma}{d\Phi}=|\sum A(J^{PC})|^{2},
\end{equation}
where $A(J^{PC})$ is the full amplitude for all possible resonances whose
spin-parity are $J^{PC}$. $\int
d\xi\omega(\xi)\epsilon(\xi)\equiv\sigma'$ is the measured total cross
section.

The joint probability density for observing the $N$ events in the data
sample is:
\begin{equation}
\el=\prod\limits_{i=1}^{N}P(\xi_{i})=\prod\limits_{i=1}^{N}\frac{(\frac{d\sigma}{d\Phi})_{i}\epsilon(\xi_{i})}{\sigma'}.
\end{equation}

For technical reasons, rather than maximizing $\el$, ${\cal S} = -\ln {\cal L}$ is minimized, with
\begin{eqnarray}
{\cal S} = - \ln {\cal L} = - \sum_{i=1}^{N}\ln \left(\frac{(\frac{d \sigma}{d \Phi})_i}{\sigma'}\right )-\sum_{i=1}^{N}\ln\epsilon(\xi_i),
\end{eqnarray}
for a given data set. The second term is a constant and has no impact
on the determination of the parameters of the amplitudes or on the
relative changes of $\cal S$ values. In the fitting, $-\ln\el$ is defined
as:
\begin{equation}
-\ln\el=-\sum_{i=1}^{N} \ln\left(\frac{(\frac{d\sigma}{d\Phi})_{i}}{\sigma'}\right) = -\sum_{i=1}^{N}\ln{\left(\frac{d\sigma}{d\Phi}\right)_{i}}+ N\ln{\sigma'}.
\end{equation}
The free parameters are optimized by
MINUIT~\cite{bibminuit}. The measured total cross section $\sigma'$
is evaluated using MC techniques. An MC sample of $N_{\rm gen}$
is generated with signal events that are distributed uniformly in phase space. These events are subjected to the selection criteria and yield
a sample of $N_{\rm acc}$ accepted events. The normalization integral is
computed as:
\begin{equation}
\int {d\xi}\omega(\xi)\epsilon(\xi)=\sigma'\to\frac{1}{N_{\rm acc}}\sum_{k}^{N_{\rm acc}}\left(\frac{d\sigma}{d\Phi}\right)_{k}.
\end{equation}

Since data contains the contribution of signal and background,
the contribution of non-$\phi \phi$ background events is taken into account by subtracting the
negative log-likelihood (NLL) value obtained for events in the $\phi \phi$ sidebands from
the NLL value obtained for events in the $\phi \phi$ signal region, i.e.,
\begin{equation}
\el_{\rm sig}=\frac{\el_{\rm data}}{\el_{\rm bkg}},
\end{equation}
\begin{equation}
-\ln \el_{\rm sig}=-(\ln \el_{\rm data} - \ln \el_{\rm bkg}).
\end{equation}

The number of the fitted events $N_{X}$ for an intermediate resonance
$X$, which has $N_{W_{X}}$ independent partial wave amplitudes $A_{i}$, is
defined as
\begin{equation}
 N_{X}=\frac{\sigma_{X}}{\sigma'}\cdot N',
\end{equation}
where $N^{'}$ is the number of selected events after background
subtraction, and
\begin{equation}
\sigma_{X}=\frac{1}{N_{\rm acc}}\sum_{k}^{N_{\rm acc}}|\sum_{j}^{N_{W_{X}}}(A_{j})_{k}|^{2},
\end{equation}
is the measured cross section of the resonance $X$ and is calculated
with the same MC sample as the measured total cross section $\sigma'$.

The branching fraction of $J/\psi\to\gamma X, X \to\phi\phi$ is calculated as:
\begin {equation}
\br(J/\psi\rightarrow\gamma X\rightarrow\gamma\phi\phi)=\frac{N_{X}}{N_{J/\psi}\cdot\varepsilon_{X}
\cdot \br^{2}_{\phi\rightarrow K^{+} K^{-}}},
\end {equation}
where the detection efficiency $\varepsilon_{X}$ is obtained by
the partial wave amplitude weighted MC sample,
\begin{equation}
\varepsilon_{X}=\frac{\sigma_{X}}{\sigma_{X}^{\rm gen}}=\frac{\sum_{k}^{N_{\rm acc}}|\sum_{j}^{N_{W_{X}}}(A_{j})_{k}|^{2}}{\sum_{i}^{N_{\rm gen}}|\sum_{j}^{N_{W_{X}}}(A_{j})_{i}|^{2}},
\end{equation}
$N_{J/\psi}$ is the total number of
$J/\psi$ events, and ${\cal B}_{\phi\rightarrow K^{+}K^{-}}= (48.9 \pm 0.5)\%$ is the branching fraction of
$\phi\rightarrow K^{+}K^{-}$ taken from Ref.~\cite{bibpdg}.

\subsection{PWA results}
\begin{figure*}[!htbp]
  \centering
  \includegraphics[width=0.32\textwidth,height=0.19\textheight]{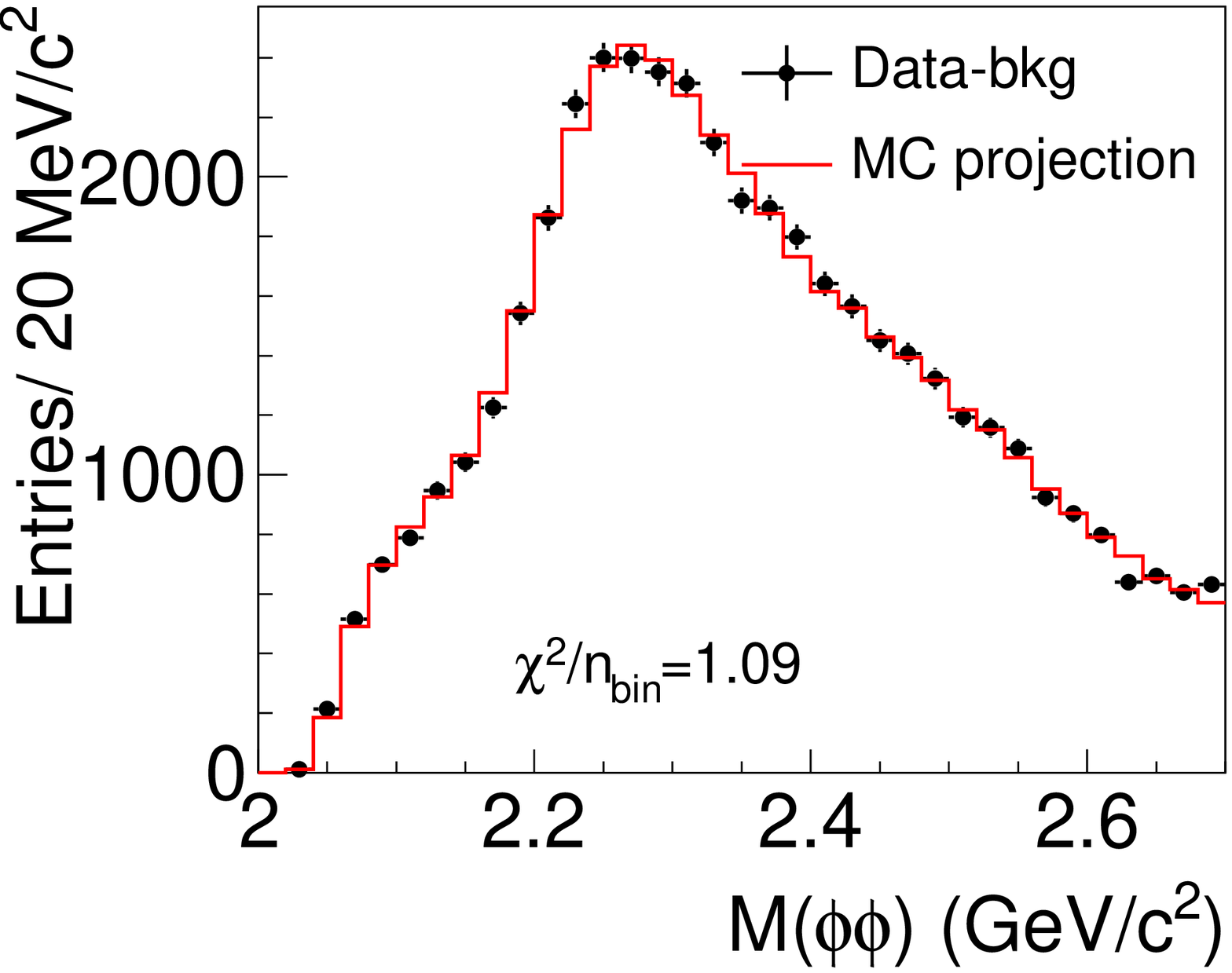}
  \put(-135,6){(a)}
  \includegraphics[width=0.32\textwidth,height=0.19\textheight]{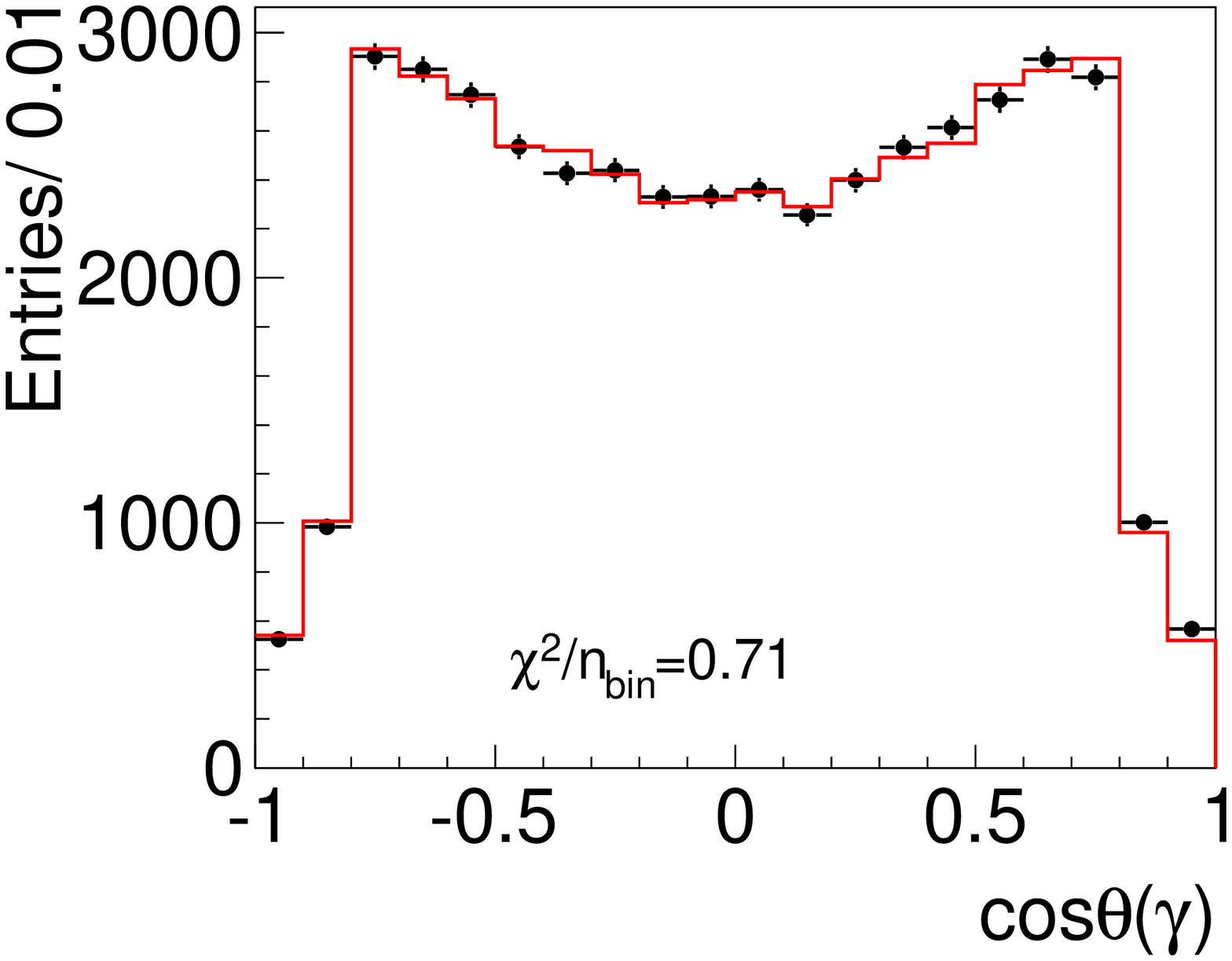}
  \put(-135,6){(b)}
  \includegraphics[width=0.32\textwidth,height=0.19\textheight]{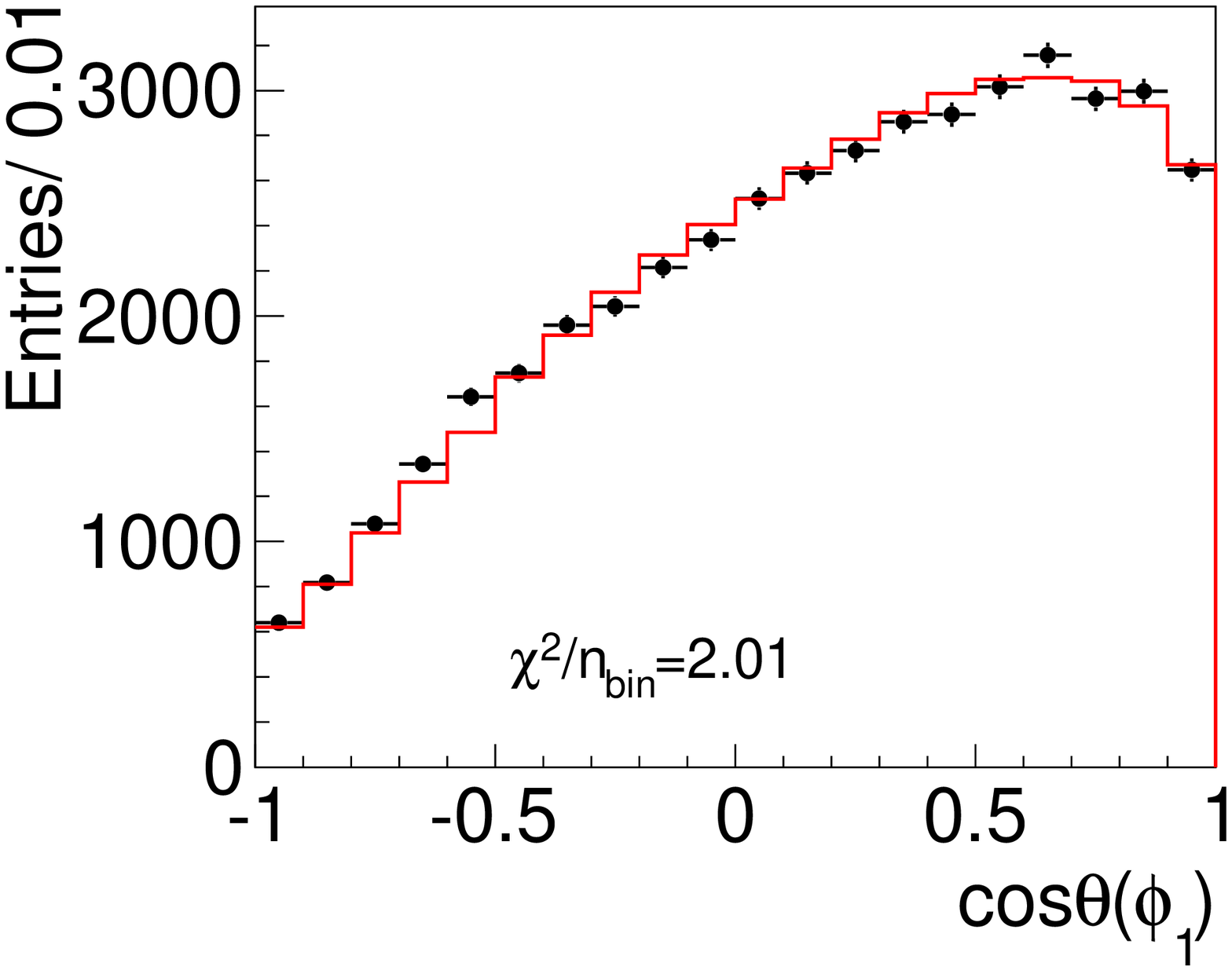}
  \put(-135,6){(c)} \\
  \includegraphics[width=0.32\textwidth,height=0.19\textheight]{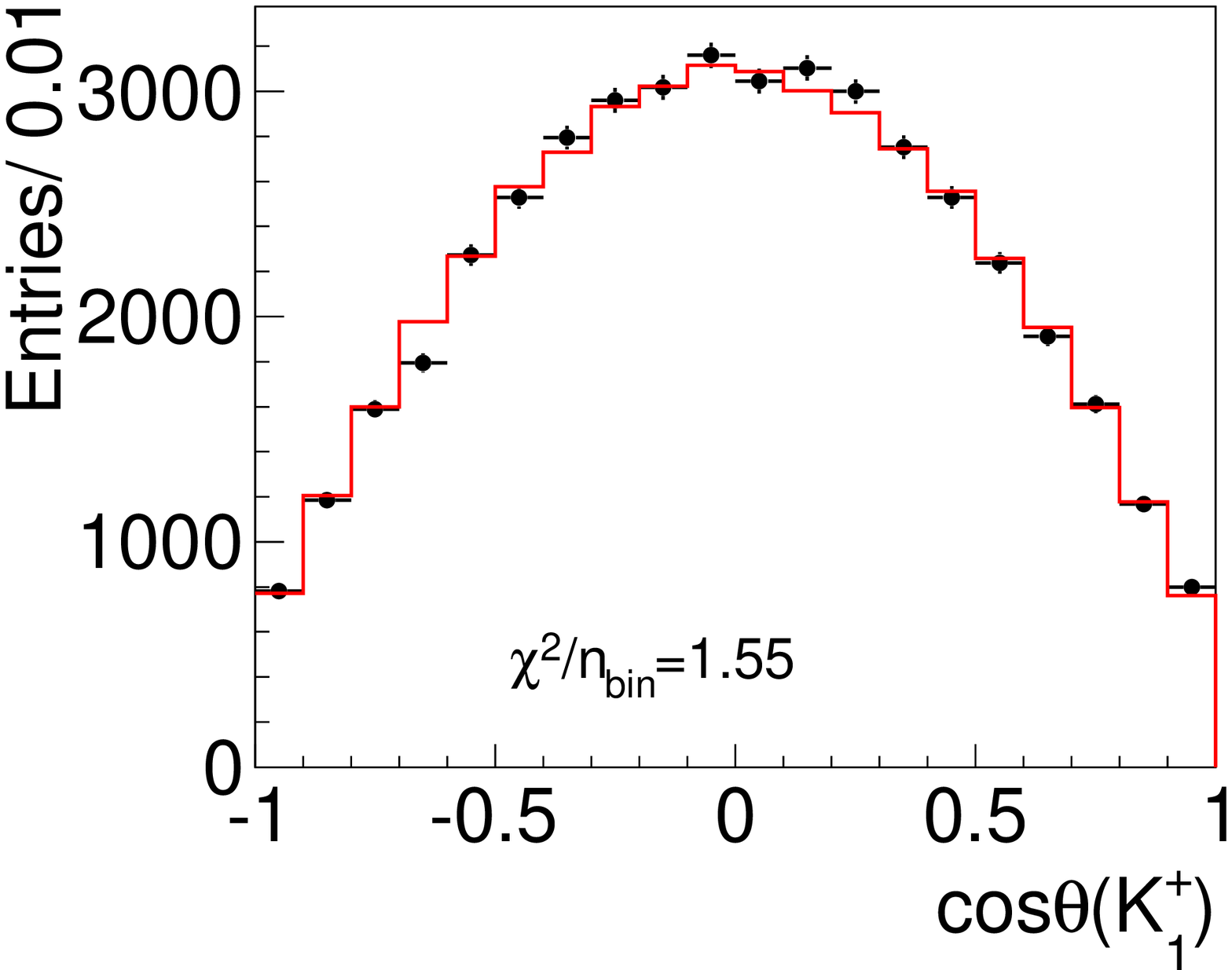}
  \put(-135,6){(d)}
  \includegraphics[width=0.32\textwidth,height=0.19\textheight]{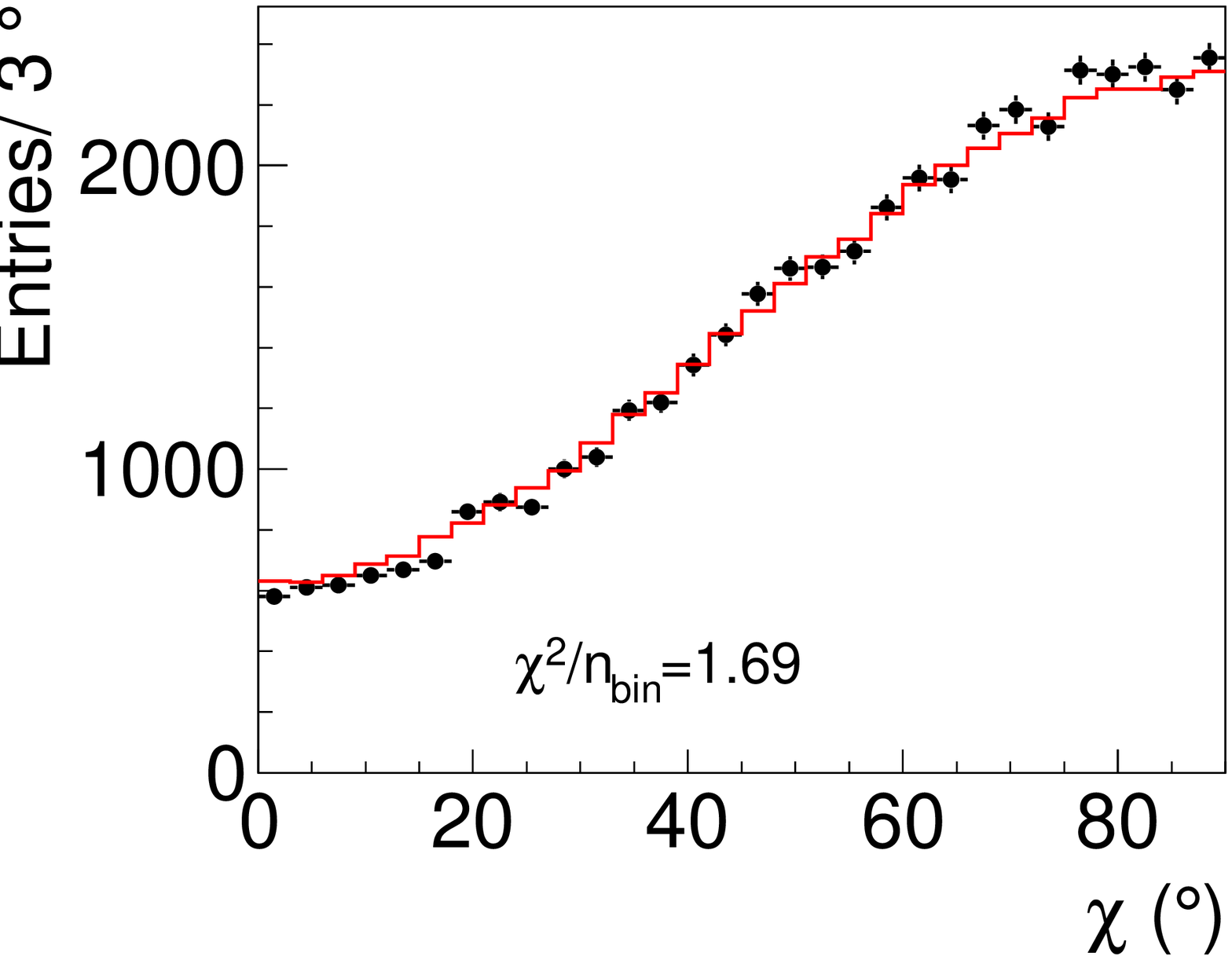}
  \put(-135,6){(e)}
  \includegraphics[width=0.32\textwidth,height=0.19\textheight]{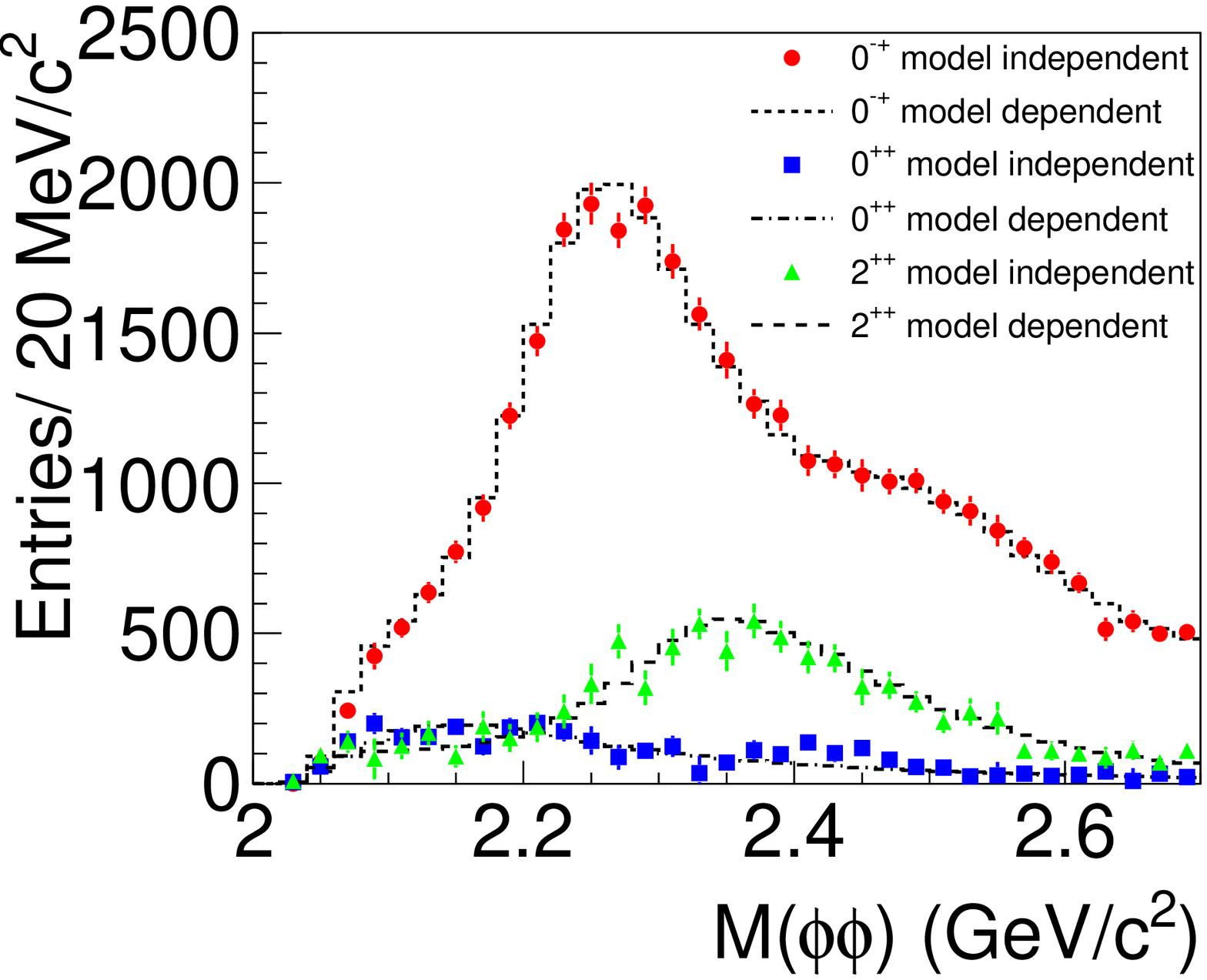}
  \put(-135,6){(f)}
  \caption{Superposition of data and the PWA fit projections for: (a) invariant mass
    distributions of $\phi\phi$; (b) $\cos\theta$ of $\gamma$ in the $J/\psi$ rest frame;
    (c) $\cos\theta$ of $\phi_1$ in the $X$ rest frame; (d) $\cos\theta$ of $K^{+}$ in the
    $\phi_1$ rest frame; (e) the azimuthal angle between the normals to the two decay
    planes of $\phi$ in the $X$ rest frame. Black dots with error bars are data with
    background events subtracted and the solid red lines are projections of the model-dependent fit. (f) Intensities of individual $J^{PC}$ components. The red dots, blue
    boxes and green triangles with error bars are the intensities of $J^{PC} = 0^{-+}$,
    $0^{++}$ and $2^{++}$, respectively, from the model-independent fit in each bin. The
    short-dashed, dash-dotted and long-dashed histograms show the coherent superpositions
    of the BW resonances with $J^{PC} = 0^{-+}$, $0^{++}$ and $2^{++}$, respectively,
    from the model-dependent fit.}
  \label{fig:projection}
\end{figure*}

In this analysis, all possible combinations of $J^{PC}$ = $0^{-+}$, $0^{++}$ and $2^{++}$
resonances~\cite{bibreslist} listed in the PDG~\cite{bibpdg} are evaluated. Given the small phase space of $J/\psi\rightarrow \gamma \phi \phi$, $J\geq4$ states should be suppressed. The changes
in the NLL value and the number of free parameters in the fit with and without a resonance
are used to evaluate its statistical significance. In the baseline solution, there are
three $0^{-+}$ resonances ($\eta(2225)$, $\eta(2100)$, and $X(2500)$), one $0^{++}$ resonance
($f_0(2100)$), three $2^{++}$ resonances ($f_2(2010)$, $f_2(2300)$, and $f_2(2340)$), and
the direct decay of $J/\psi\rightarrow \gamma \phi \phi$, which is modeled by a $0^{-+}$
phase space distribution ($0^{-+}$ PHSP) of the $\phi \phi$ system. The statistical significance of
each component in the baseline solution is larger than 5~$\sigma$. The masses and widths
of the three $0^{-+}$ resonances are free parameters in the fit. The resonance parameters
of the $0^{++}$ and $2^{++}$ resonances are fixed to the PDG~\cite{bibpdg} values due to
limited statistics. The masses and widths of the resonances, product branching fractions
of $J/\psi\rightarrow \gamma X$, $X \rightarrow \phi\phi$, and the statistical
significances are summarized in Table~\ref{tab:pwasolution}, where the first errors are
statistical, and the second ones are systematic. The fit fraction of each component and
their interference fractions are shown in Table~\ref{tab:fraction}.
Figure~\ref{fig:projection}(a)
shows a comparison of the data and the PWA fit projection (weighted by MC efficiencies) of
the invariant mass distributions of $\phi\phi$ for the fitted parameters. The comparisons
of the projected data and MC angular distributions for the events with $\phi\phi$
invariant mass less than 2.7 GeV$/c^{2}$ are shown in
Fig.~\ref{fig:projection}(b)$-$\ref{fig:projection}(e).
The $\chi^2/n_{\rm bin}$ value is displayed on each figure to demonstrate the goodness of fit,
where $n_{\rm bin}$ is the number of bins of each figure and $\chi^2$ is defined as:
\begin{eqnarray}
\chi^2=\sum_{i=1}^{n_{\rm bin}}\frac{(n_i-\nu_i)^2}{\nu_i},
\end{eqnarray}
where $n_i$ and $\nu_i$ are the number of events for the data and the
fit projections with the baseline solution in the $i$th bin of each
figure, respectively.

\begin{table}[!htbp]
  \centering
  \caption{\label{tab:pwasolution} Mass, width, ${\cal B}(J/\psi\rightarrow \gamma X
    \rightarrow \gamma \phi\phi)$ (B.F.) and significance (Sig.) of each component in the
    baseline solution. The first errors are statistical and the second ones are
    systematic.}
  \linespread{1.5}
  \begin{small}
    \resizebox{0.48\textwidth}{!}{%
      \begin{tabular}{ccccc}
        \hline
        \hline
        Resonance     &M(MeV/$c^{2}$) &$\Gamma$(MeV/$c^{2}$) &B.F.($ \times 10^{-4}$)  &Sig.\\
        \hline
        $\eta(2225)$  &$2216_{-5}^{+4}$$_{-11}^{+21}$      &$185_{-14}^{+12}$$_{-17}^{+43}$    &$(2.40\pm0.10$$_{-0.18}^{+2.47})$         &28$\;\sigma$        \\
        $\eta(2100)$  &$2050_{-24}^{+30}$$_{-26}^{+75}$     &$250_{-30}^{+36}$$_{-164}^{+181}$ &$(3.30\pm0.09$$_{-3.04}^{+0.18})$         &22$\;\sigma$      \\
        $X(2500)$       &$2470_{-19}^{+15}$$_{-23}^{+101}$      &$230_{-35}^{+64}$$_{-33}^{+56}$  &$(0.17\pm0.02$$_{-0.08}^{+0.02})$         &8.8$\;\sigma$\\
        $f_{0}(2100)$ &2101                     &224                                           &$(0.43\pm0.04$$_{-0.03}^{+0.24})$         &24$\;\sigma$\\
        $f_{2}(2010)$ &2011                     &202                                           &$(0.35\pm0.05$$_{-0.15}^{+0.28})$         &9.5$\;\sigma$ \\
        $f_{2}(2300)$ &2297                     &149                                           &$(0.44\pm0.07$$_{-0.15}^{+0.09})$         &6.4$\;\sigma$ \\
        $f_{2}(2340)$ &2339                     &319                                           &$(1.91\pm0.14$$_{-0.73}^{+0.72})$         &11$\;\sigma$\\
        $0^{-+}$ PHSP &                         &                                              &$(2.74\pm0.15$$_{-1.48}^{+0.16})$         &6.8$\;\sigma$\\
        \hline
        \hline
      \end{tabular}
    }%
  \end{small}
\end{table}

\begin{table*}[!htbp]
\centering
\linespread{1.5}
\begin{small}
\caption{\label{tab:fraction} Fraction of each component and interference fractions between two components (\%) in the baseline solution. The errors are statistical only.}
\begin{tabular}{ccccccccc}
\hline\hline
Resonance      &$\eta(2100)$     &$\eta(2225)$   &$X(2500)$ &$0^{-+}$ PHSP &$f_{0}(2100)$  &$f_{2}(2010)$ &$f_{2}(2300)$ &$f_{2}(2340)$ \\
\hline
$\eta(2100)$     &54.2$\pm$1.5  &43.5$\pm$1.2   &15.2$\pm$1.0   &$-$64.0$\pm$2.2        &0.0$\pm$0.0    &0.0$\pm$0.0    &0.0$\pm$0.0    &$-$0.1$\pm$0.0 \\
$\eta(2225)$     & &41.0$\pm$1.6        &15.9$\pm$0.7   &$-$60.6$\pm$1.7        &0.0$\pm$0.0    &0.0$\pm$0.0    &0.1$\pm$0.0    &$-$0.1$\pm$0.0 \\
$X(2500)$        & & &3.2$\pm$0.3       &$-$15.7$\pm$1.0        &0.0$\pm$0.0    &0.0$\pm$0.0    &0.0$\pm$0.0    &0.0$\pm$0.0\\
$0^{-+}$ PHSP    & & & &42.8$\pm$2.3    &0.0$\pm$0.0    &0.0$\pm$0.0    &0.0$\pm$0.0    &0.0$\pm$0.0 \\
$f_{0}(2100)$    & & & & &6.5$\pm$0.6   &0.1$\pm$0.0    &0.1$\pm$0.0    &$-$0.5$\pm$0.0\\
$f_{2}(2010)$    & & & & & &5.9$\pm$0.8 &6.0$\pm$0.7    &$-$18.6$\pm$1.6\\
$f_{2}(2300)$    & & & & & & &8.8$\pm$1.4       &$-$22.0$\pm$3.5\\
$f_{2}(2340)$    & & & & & & & &38.4$\pm$2.8\\
\hline\hline
\end{tabular}
\end{small}
\end{table*}

Various checks are performed to test the reliability of the model-dependent PWA
solution. Replacing the pseudoscalar state $\eta(2100)$ by either
$\eta(2010)$~\cite{bib2010} or $\eta(2320)$~\cite{bib2320} worsens the NLL values by 21.2
and 33.0, respectively. The spin-parity assignment $J^{PC}$ of the
$X(2500)$ as $0^{-+}$ is significantly better than the $0^{++}$
hypothesis, with the NLL value improving by 44.1 units.  Changing the
spin-parity assignment of the $X(2500)$ to $2^{++}$, resulting in 10
additional free parameters, worsens the NLL value by 0.5, instead. Therefore, the
preferred assignment for the $X(2500)$ is pseudoscalar. If we replace the two tensor states
$f_2(2300)$ and $f_2(2340)$ by a single one with free resonance parameters in the fit, the
NLL value is worsened by 14.7. In this case, a statistical significance test of the
$f_2(2340)$ yields a value of $6.1\;\sigma$. The narrow $f_J(2220)$ (alternatively known as
the $\xi(2230)$), which was seen in $J/\psi\rightarrow \gamma K^+ K^-$ at MarkIII~\cite{bib2220mark3} and
BES~\cite{bib2220bes}, but not seen in $J/\psi\rightarrow \gamma K_S^0 K_S^0$ at CLEO~\cite{bib2220cleo}, is also studied. When included in the PWA, the statistical significance of the $f_J(2220)$ is found to be $0.8\;\sigma$. The upper limit on the branching fraction ratio ${\cal B}(\xi(2230)\rightarrow \phi\phi)/{\cal B}(\xi(2230)\rightarrow K^+K^-)$ at the 90\% C.L. is estimated to be $1.91\times10^{-2}$. For the description of the nonresonant contribution, the statistical significance of additional non-resonant
contributions with $J^{PC}$ = $0^{++}$ or $2^{++}$ is less than $5\;\sigma$. Additional
resonances listed in Ref.~\cite{bibpdg} as well as two extra states, the $X(2120)$ and
$X(2370)$ from Ref.~\cite{bib2370}, are tested with all possible $J^{PC}$ assignments. None
of them has a statistical significance larger than $5\;\sigma$, as
shown in Table~\ref{tab:extra_res}. The existence of possible additional
resonances is further studied by performing scans for extra resonances ($J^{PC}$ =
$0^{-+}$, $0^{++}$, $1^{++}$, $2^{-+}$, $2^{++}$ and $4^{++}$) with different masses and
widths. The scan results yield no evidence for extra intermediate states.  The reliability
of the fit procedure is tested by an input-output check, as follows:  An MC sample is generated
with given components. After the fitting procedure described above, the properties of the
components (mass, width, branching fraction, and the effect of interference terms) are
compared with the input values. The output values agree with the input around
$\pm 1\;\sigma$, confirming the reliability of the fitting procedure.

\begin{table}[!htbp]
\centering
\linespread{1.5}
\begin{small}
\caption{\label{tab:extra_res} Additional resonances, $J^{PC}$, change of number of free parameters ($\Delta$Ndof), change of NLL ($\Delta$NLL ) and corresponding significance (Sig.).}
\begin{tabular}{ccccc}
\hline
\hline
Resonance                                          &$J^{PC}$         &$\Delta$Ndof &$\Delta$NLL &Sig. \\
\hline
$f_{0}(2020)$                                      &$0^{++}$        &4     &11.5    &3.8$\;\sigma$\\
$f_{0}(2330)$                                      &$0^{++}$        &4     &4.3     &1.8$\;\sigma$\\
$f_{0}(2200)$                                      &$0^{++}$        &4     &5.0     &2.0$\;\sigma$\\
$f_{2}(2150)$                                      &$2^{++}$        &12    &25.1    &4.8$\;\sigma$\\
$f_{J}(2220)$                                      &$2^{++}$        &12    &6.3     &0.8$\;\sigma$\\
$\eta(2010)$                                       &$0^{-+}$        &2     &1.5     &1.2$\;\sigma$\\
$\eta(2320)$                                       &$0^{-+}$        &2     &0.4     &0.4$\;\sigma$\\
\multicolumn{1}{c|}{\multirow{3}{*}{$X(2370)$}}    &$0^{-+}$        &2     &0.5     &0.5$\;\sigma$\\
                                                   &$0^{++}$        &4     &5.4     &2.2$\;\sigma$\\
                                                   &$2^{++}$        &12    &17.8    &3.5$\;\sigma$\\
\multicolumn{1}{c|}{\multirow{3}{*}{$X(2120)$}}      &$0^{-+}$        &2     &1.3     &1.1$\;\sigma$\\
                                                   &$0^{++}$        &4     &2.3     &0.9$\;\sigma$\\
                                                   &$2^{++}$        &12    &14.9    &3.0$\;\sigma$\\
\hline
\hline
\end{tabular}
\end{small}
\end{table}

In addition to the PWA fit with resonances described by BW functions, a model-independent
fit where the intermediate states are parameterized by a separate complex constant for each of 35 bins of 20 MeV/$c^2$ width is performed in the region $M(\phi\phi) < 2.7$
GeV/$c^{2}$ to extract the contribution of components with each $J^{PC}$ using the method described in Ref.~\cite{bibmodelindep}. The fit results are
shown in Fig.~\ref{fig:projection}(f). The $0^{-+}$ contribution is dominant, and a
strong $2^{++}$ component at 2.3 GeV/$c^2$ is observed. In general, the model-independent fit
gives similar features to those of the model-dependent fit, and the results of these two
fits are consistent with each other.

\section{Systematic uncertainties}
The sources of systematic uncertainty are divided into two categories. The first includes
the systematic uncertainties from the number of $J/\psi$ events (0.8\%~\cite{numsys1,numsys2}), MDC
tracking (1.0\% each for three charged tracks~\cite{photonsys}), kaon PID (1.0\% each for
three kaons~\cite{photonsys}), photon detection efficiency (1.0\%~\cite{photonsys}),
kinematic fit (2.5\%), $\phi$ mass resolution (0.3\%) and
${\cal B}_{\phi\rightarrow K^{+}K^{-}}$ (2.0\%). These systematic uncertainties are
applicable to all the branching fraction measurements. The total systematic uncertainty
from these sources is 5.5\%. The second source concerns the PWA fit procedure, where
the systematic uncertainties are applicable to measurements of the branching fractions and
resonance parameters. These sources of systematic uncertainties are described below.
\begin{itemize}
\item[(i)] BW parametrization. Uncertainties from the BW parametrization are estimated by the
changes in the fit results caused by replacing the fixed width $\Gamma_0$ of the BW for
the threshold states $\eta(2100)$ and $\eta(2225)$ with a mass-dependent width form
$\Gamma(m)$~\cite{bibmdbw}.

\item[(ii)] Uncertainty from resonance parameters. In the nominal fit, the resonance parameters of the $0^{++}$
and $2^{++}$ states are fixed. An alternative fit is performed in which those resonance
parameters are varied within one standard deviation of the PDG values~\cite{bibpdg}, and
the changes in the results are taken as systematic uncertainties.

\item[(iii)] Background uncertainty. To estimate the background uncertainty, alternative fits are
performed with background events from different $\phi$ sideband regions and different
normalization factors, and the changes in the results are assigned as the systematic
uncertainties.

\item[(iv)] Uncertainty from additional resonances. Uncertainties from possible additional resonances are estimated by adding the $f_0(2020)$
and the $f_2(2150)$, which are the two most significant additional resonances, into the
baseline configuration individually, the changes of the measurements caused by them are assigned as the systematic uncertainties.
\end{itemize}
For each alternative fit performed to estimate the systematic uncertainties from the PWA fit procedure, the changes of the measurements are taken as the one-sided systematic uncertainties.
For each measurement, the individual uncertainties are assumed to be
independent and are added in quadrature to obtain the total systematic uncertainty on the negative and positive side, respectively.
The sources of systematic uncertainties applicable to the measurements of masses and widths of $\eta(2225)$, $\eta(2100)$ and $X(2500)$, and their contributions are summarized in Table~\ref{tab:sys_masswidth}. The relative systematic uncertainties relevant for the branching fraction measurements are summarized in Table~\ref{tab:sys_br}, where the last row is the total relative systematic uncertainty from fitting irrelevant sources.

\begin{table*}[!htbp]
\centering
\linespread{1.5}
\caption{\label{tab:sys_masswidth} Summary of the systematic error sources and their corresponding contributions (in MeV/$c^{2}$) to the systematic uncertainties in masses and widths of $\eta(2100)$, $\eta(2225)$ and $X(2500)$, denoted as $\Delta$M and $\Delta \Gamma$, respectively.}
\begin{tabular}{cccccccc}
\hline
\hline
\multicolumn{2}{c}{\multirow{2}{*}{Sources}} &\multicolumn{2}{c}{$\eta(2100)$} &\multicolumn{2}{c}{$\eta(2225)$}  &\multicolumn{2}{c}{$X(2500)$} \ST \\\cline{3-8}
\multicolumn{2}{c}{}                        &$\Delta$M &$\Delta \Gamma$ &$\Delta$M &$\Delta \Gamma$ &$\Delta$M &$\Delta \Gamma$ \ST \\
\hline
\multicolumn{2}{c}{Breit-Wigner parametrization}      &$_{-10}^{+72}$  &$_{-152}^{+164}$  &$_{-10}^{+9}$   &$_{-0}^{+43}$  &$_{-5}^{+20}$ &$_{-30}^{+15}$\ST \\
\multicolumn{2}{c}{Resonance parameters}     &$_{-0}^{+1}$    &$_{-0}^{+1}$      &$_{-1}^{+0}$    &$_{-1}^{+0}$   &$_{-2}^{+0}$  &$_{-3}^{+0}$  \ST \\
\multicolumn{2}{c}{Background uncertainty}    &$_{-22}^{+20}$  &$_{-10}^{+64}$    &$_{-5}^{+11}$   &$_{-5}^{+6}$  &$_{-20}^{+42}$ &$_{-10}^{+36}$  \ST \\
\multicolumn{1}{c}{\multirow{2}{*}{Extra resonances}} &$f_2(2150)$   &$_{-10}^{+0}$  &$_{-0}^{+40}$  &$_{-0}^{+10}$ &$_{-6}^{+0}$   &$_{-10}^{+0}$   &$_{-10}^{+0}$\ST \\ 
                                  &other insignificant resonances     &$_{-0}^{+10}$  &$_{-60}^{+0}$  &$_{-0}^{+12}$ &$_{-15}^{+0}$  &$_{-0}^{+90}$ &$_{-0}^{+40}$ \ST \\
\multicolumn{2}{c}{Total}                  &$_{-26}^{+75}$  &$_{-164}^{+181}$ &$_{-11}^{+21}$ &$_{-17}^{+43}$ &$_{-23}^{+101}$  &$_{-33}^{+56}$  \ST \\
\hline
\hline
\end{tabular}
\end{table*}

\begin{table*}[!htbp]
\centering
  \linespread{1.5}
\caption{\label{tab:sys_br} Summary of the systematic error sources and their corresponding contributions to the branching fractions of $J/\psi\rightarrow \gamma X \rightarrow \gamma \phi\phi$ (relative uncertainties, in \%), which are denoted as $\Delta {\cal B}$.}
\begin{tabular}{cccccccccc}
\hline
\hline
\multicolumn{2}{c}{Sources}     &$\eta(2100)$ &$\eta(2225)$  &$X(2500)$  &$f_{0}(2100)$ &$f_{2}(2010)$ &$f_{2}(2300)$ &$f_{2}(2340)$ &$0^{-+}$ PHSP   \ST   \\
\hline
\multicolumn{2}{c}{Event selection}       &$\pm5.5$   &$\pm5.5$  &$\pm5.5$ &$\pm5.5$ &$\pm5.5$ &$\pm5.5$ &$\pm5.5$ &$\pm5.5$                \ST \\
\multicolumn{2}{c}{Breit-Wigner parametrization}  &$_{-91.8}^{+0.0}$ &$_{-0.0}^{+102.9}$ &$_{-48.0}^{+0.0}$ &$_{-2.1}^{+0.4}$ &$_{-0.0}^{+23.7}$ &$_{-0.6}^{+7.9}$  &$_{-12.3}^{+0.0}$ &$_{-53.4}^{+0.0}$       \ST \\
\multicolumn{2}{c}{Resonance parameters}  &$_{-2.7}^{+0.0}$ &$_{-3.7}^{+0.0}$ &$_{-11.8}^{+0.0}$ &$_{-0.0}^{+0.9}$ &$_{-11.9}^{+0.0}$ &$_{-0.0}^{+15.7}$  &$_{-13.7}^{+0.0}$ &$_{-5.8}^{+0.0}$       \ST \\
\multicolumn{2}{c}{Background uncertainty} &$_{-0.2}^{+0.7}$  &$_{-0.1}^{+0.9}$  &$_{-0.1}^{+10.4}$ &$_{-0.1}^{+1.8}$ &$_{-3.1}^{+1.6}$ &$_{-3.3}^{+7.4}$  &$_{-0.7}^{+1.2}$ &$_{-0.2}^{+1.7}$  \ST \\
\multicolumn{1}{c}{\multirow{2}{*}{Extra resonances}} &$f_2(2150)$ &$_{-3.1}^{+0.0}$  &$_{-3.4}^{+0.0}$  &$_{-4.5}^{+0.0}$ &$_{-0.1}^{+0.0}$ &$_{-0.0}^{+75.6}$ &$_{-18.1}^{+0.0}$  &$_{-0.0}^{+37.3}$ &$_{-3.0}^{+0.0}$ \ST \\ 
&other insignificant resonances                                     &$_{-0.6}^{+0.0}$ &$_{-2.0}^{+0.0}$ &$_{-0.8}^{+0.0}$ &$_{-0.0}^{+56.6}$ &$_{-41.3}^{+0.0}$&$_{-28.1}^{+0.0}$ &$_{-32.8}^{+0.0}$ &$_{-2.8}^{+0.0}$ \ST \\

\multicolumn{2}{c}{Total}              &$_{-92.1}^{+5.5}$  &$_{-7.7}^{+103.1}$ &$_{-49.9}^{+11.8}$ &$_{-5.9}^{+56.9}$ &$_{-43.4}^{+79.4}$   &$_{-34.0}^{+19.8}$  &$_{-38.0}^{+37.7}$  &$_{-54.2}^{+5.8}$\ST \\
\hline
\hline
\end{tabular}
\end{table*}

\section{Summary}
In summary, a PWA on $J/\psi\rightarrow\gamma\phi\phi$ has been performed
based on (1310.6 $\pm$ 10.5) $\times 10^{6}$ $J/\psi$ events collected
with the BESIII detector.
The most remarkable feature of the PWA results is
that $0^{-+}$ states are dominant. The existence of the $\eta(2225)$
is confirmed and two additional pseudoscalar states, $\eta(2100)$ with a
mass $2050_{-24}^{+30}$$_{-26}^{+75}$ MeV/$c^{2}$ and
a width $250_{-30}^{+36}$$_{-164}^{+181}$ MeV/$c^{2}$
and $X(2500)$ with a mass $2470_{-19}^{+15}$$_{-23}^{+101}$ MeV/$c^{2}$ and
a width $230_{-35}^{+64}$$_{-33}^{+56}$ MeV/$c^{2}$, are observed.
The new experimental results are helpful for mapping out pseudoscalar
excitations and searching for a $0^{-+}$ glueball.
The three tensors $f_2(2010)$, $f_2(2300)$ and $f_2(2340)$ observed in
$\pi^- p\rightarrow \phi\phi n$~\cite{bibpiN} are also observed
in $J/\psi\rightarrow\gamma\phi\phi$. Recently, the production rate
of the pure gauge tensor glueball in $J/\psi$ radiative decays has been
predicted by Lattice QCD~\cite{bibtensorglueball}, which is compatible
with the large production rate of the $f_2(2340)$
in $J/\psi\rightarrow\gamma\phi\phi$ and $J/\psi\rightarrow\gamma\eta\eta$~\cite{bibgee}.

\section*{Acknowledgements}
The BESIII Collaboration thanks the staff of BEPCII and
the IHEP computing center for their strong support. This
work is supported in part by the National Key Basic
Research Program of China under Contract
No. 2015CB856700; National Natural Science
Foundation of China (NSFC) under Contracts
No. 11235011, No. 11322544, No. 11335008,
No. 11425524; the Chinese Academy of Sciences (CAS)
Large-Scale Scientific Facility Program; the CAS Center
for Excellence in Particle Physics (CCEPP); the
Collaborative Innovation Center for Particles and
Interactions (CICPI); Joint Large-Scale Scientific Facility
Funds of the NSFC and CAS under Contracts
No. U1232201, No. U1332201; CAS under Contracts
No. KJCX2-YW-N29, No. KJCX2-YW-N45; 100
Talents Program of CAS; National 1000 Talents Program
of China; INPAC and Shanghai Key Laboratory for Particle
Physics and Cosmology; Istituto Nazionale di Fisica
Nucleare, Italy; Joint Large-Scale Scientific Facility
Funds of the NSFC and CAS under Contract
No. U1532257; Joint Large-Scale Scientific Facility
Funds of the NSFC and CAS under Contract
No. U1532258; Koninklijke Nederlandse Akademie van
Wetenschappen (KNAW) under Contract No. 530-
4CDP03; Ministry of Development of Turkey under
Contract No. DPT2006K-120470; The Swedish Resarch
Council; U.S. Department of Energy under Contracts
No. DE-FG02-05ER41374, No. DE-SC-0010504,
No. DE-SC0012069, No. DESC0010118; U.S. National
Science Foundation; University of Groningen (RuG) and
the Helmholtzzentrum fuer Schwerionenforschung GmbH
(GSI), Darmstadt; WCU Program of National Research
Foundation of Korea under Contract No. R32-2008-000-
10155-0.

\end{document}